# Challenges of Upgrading a Virtual Appliance

Kamran Karimi
Burnaby, BC, Canada
kamran@kamran-karimi.com

**Abstract**

A virtual appliance contains a target application, and the running environment necessary for running that application. Users run an appliance using a virtualization engine, freeing them from the need to make sure that the target application has access to all its dependencies. However, creating and managing a virtual appliance, versus a stand-alone application, requires special considerations. Upgrading a software system is a common requirement, and is more complicated when dealing with an appliance. This is because both the target application and the running environment must be upgraded, and there are often dependencies between these two components. In this paper we briefly discuss some important points to consider when upgrading an appliance. We then present a list of items that can help developers prevent problems during an upgrade effort.

## 1. Introduction

Virtualization is considered a good technique to solve problems such as ever-increasing physical server count and power consumption. There are many virtualization technologies and engines, such as VMware ESXi [2], Xen [4], and VirtualBox [6]. A virtual appliance is a popular method of enabling users to easily benefit from an application's functionality [8]. A major reason for producing a virtual appliance is allowing the user to take advantage of an application (called the target application here) without the need to worry about any hardware or software requirements.

The target application may be a single piece of software, or it may be very complex and comprise many subparts and rely on components such as databases and authentication services. The application, together with all its needed components, is packaged as a virtual machine. The appliance is tested and delivered to the user, who runs it using a virtualization engine. From a software engineering point of view, a virtual appliance is a convenient method of distributing an application, because it provides total control over the running environment of an application. Developers can make sure that specific software libraries or even hardware components are present. As a result, there is no need to worry about incompatible operating systems, missing or incompatible link libraries, inappropriate configuration of the system, or special hardware equipment. Users simply have to make sure that the virtualization environment they are using supports the virtual appliance.

This simplicity for the user comes at the expense of complexity at the development end. In addition to developing the application, there is now a need to put together a working system that contains all the needed components to run it. Care is usually taken to make sure that unnecessary software is not included in the appliance. This ensures that the minimum needed storage is used, attack vectors that could be exploited by hackers are minimized [3], and any conflicts that may arise in a complex setup are avoided. Having as small an appliance as possible also makes upgrading the appliance easier. In this paper we briefly discuss some of the major issues that arise when an appliance is upgraded.

The rest of the paper presents common points that may be encountered when upgrading an appliance. They are based on experience with appliances that were in wide use. Section 2 expands some of the problems of upgrading an appliance. In Section 3 we itemise some of the major points to consider during an upgrade. Section 4 concludes the paper.

## 2. Upgrade Challenges

A virtual appliance, like any other software product, may need to be upgraded to fix bugs or add features. Since an appliance is often a self-contained system, inter-dependencies between the different components often lead to complications during an upgrade. The increasing popularity of appliances is bringing attention to this problem, and efforts are underway to make an upgrade as automatic as possible [1, 5]. However, software is chaotic by nature, meaning that a change in one part of the system may cause a side-effect in another unpredictable part. For this reason, it may be a good idea to approach the upgrade as a typical software engineering problem that needs careful consideration and attention.

Upgrading a virtual appliance comprises two main activities: upgrading the application, and upgrading the running environment. These two activities are related. When an application is modified, one may decide to use a more recent programming environment. This could mean a newer version of the programming language with added features, or a new, debugged set of programming libraries. When using an interpreted programming language such as Ruby or Python, the implications of moving to a newer version of the language may be more serious. For example, the target application may need to change for deprecated language components to be removed.

Operating System (OS) versions usually have a limited support period. A few years after their release, OS vendors stop providing bug fixes or security updates, practically forcing users to upgrade to a newer version of the OS. This holds for an appliance too, and even if the target application has not changed, an obsolete OS needs to be replaced with a more recent one. This often implies upgrading the link libraries, in turn probably making it necessary to use a newer programming environment to build the application. Major changes may be needed if the programming language has undergone radical changes.

A good question to ask is whether one should start with changing the application (if any changes are needed), or with the running environment. The answer depends on the case at hand. If a change in requirements of the application started the upgrade process, then obviously the application should be modified first. Any need to change the running environment should then be noted and implemented. These changes may in turn make it necessary to change further components of the appliance, and so on.

If, on the other hand, the need to apply bug or security fixes in the running environment are causing the upgrade, then one should consider what the implications are on the application. For example, if moving to a newer OS version makes it necessary to move to a newer language environment, one may have to modify the target application to make sure obsolete or deprecated features are replaced.

## 3. Points to Consider

Given the predictable nature of most of the issues that one may face when upgrading an appliance, we can come up with a number of points that should be considered to reduce the number of potential problems, as follows.

1) Upgrading an appliance is an iterative process. Changing the target application may cause changes in the running environment, and vice versa. If different teams are working on modifying the target application and the running environment, then good communication and passing of feedback are crucial to success. Expect the list of changed software components in an appliance to gradually become stable. The rate of progress depends on the extent of the needed changes.
2) For reasons explained before, it is better to install the minimum number of software components needed. This may be hard to achieve given the often intricate web of dependencies between system packages. Developers can list the packages that are directly needed by the target application, but what do these packages depend on? Fortunately there are tools such as gdebi in Debian-based Linux distributions that can automatically resolve any unmet dependencies.
3) When referring to other resources, for example invoking interpreters, it is better to be specific. For example, if bash or python2.4 are required for an application to run correctly, then invoking /bin/bash or /usr/bin/python2.4 is better than /bin/sh or /usr/bin/python. If a specific resource is missing, the application will simply fail to start. This alerts the developers that the running environment should be verified, and prevents apparently-random problems to appear while the target application is running.

4) Since an appliance runs in a virtual environment, it is crucial to make sure that the upgraded appliance is tested under more recent versions of the target virtualization engine(s). In a paravirtualized environment [7], the need for specific (modified) versions of the kernel and related link libraries adds another layer of complexity to the upgrade effort.

5) Supporting different virtualization environments makes it easier for the users to deploy the appliance. The Open Virtualization Format (OVF) [9] is vendor-independent and thus allows the appliance to be distributed in a single file format. This reduces packaging efforts significantly. Having a single file format eases the process of upgrading the appliance because any packaging problem can be discovered and fixed just once.

6) Another issue with running an appliance with different virtualization engines arises from the fact that the running environment may differ in subtle ways. For example, storage devices may be mounted under different names as in /dev/sda versus /dev/hda. It may be possible to protect the target application in such cases. In the example just given, using Universally Unique Identifiers (UUIDs) or labels to refer to disks alleviates the problem.

7) Making the appliance completely independent of the underlying virtualization technology may not be possible. In a paravirtualized environment, one can detect the underlying virtualization technology at boot time and load the appropriate guest tools. These guest tools usually provide better performance and compatibility. The implication is that the appliance should already include a number of guest utilities and load them as appropriate. So the appliance would explicitly support a number of virtualization engines, and not others. The appliance may still be able to run under unsupported engines, but with reduced performance.

## 4. Concluding Remarks

Virtual appliances solve many physical and software engineering problems, but they present the developers with some unique problems. One of them is dealing with the often inevitable upgrading process, which is not limited to the target application only. This paper presented a number of problems that may occur when upgrading an appliance, be it the target application, or the rest of the system. Possible solutions to these problems were proposed, and a number of general guidelines for easing the upgrade process were suggested.